\documentstyle[12pt,english]{article}
\begin{document}

\begin{center}

{\bf Quantum spin liquid in  ${\bf Sr_2CuO_3}$ induced by Acoustic
Phonons }

\end{center}
\begin{center}
{\bf S.S. Aplesnin}
 \end{center}
\begin{center}
L.V. Kirenskii Institute of Physics, Siberian Branch of the
Russian Academy of Sciences, Krasnoyarsk, 660036, Russia , FAX 07
3912 438923, E-mail : apl@iph.krasn.ru
\end{center}

We have carried out theoretical studies of the phonons and spinons
excitation spectra of $Sr_2CuO_3 $ in term of a one-dimensional
model with non-adiabatic interaction magnetic and elastic
subsystems using continuous time Monte Carlo method. Two gaps in
triplet excitation spectrum are found. The  bond defects
concentration estimated is well agreement with the experimental
data. The formation of magnetic moment on site at low temperature
$T_N \sim 5 K$ is caused by freezing quantum spin liquid with
correlation radius of antiferromagnetic ordering $\xi/a \sim 300 -
500$.


{\bf 1. Introduction}

 Quasi-one-dimensional  spin systems with antiferromagnetic
interactions have received considerable attention because of
 a number of basic physical phenomena in low dimensions, such as the
existence of a spin gap depending on the spin value or the
frustration antiferromagnetic exchange, spin Peierls transitions,
accompanied by the structural symmetry breaking. Orthorhombic
compound $Sr_2CuO_3$ is considered as quasi-1D antiferromagnetic
system. However, Raman studies reported that extra peaks were
observed along the Cu-O chain directions \cite{Abrashev}  which
implies that there might be some structural instabilities. This
effect is confirmed by polarization-dependent infrared-active
phonon spectra which have shown splitting of the stretching mode
in the Cu-O chain direction about $20 cm^{-1}$ and became evident
as T decreases \cite{Lee} . These unusual behavior are not
expected by the previous theoretical calculations.

Another unclear behavior is the uniform spin susceptibility
$\chi'(q=0)$ have decreased steeply below $T \sim 33 K$ without
the signature of three -dimensional short range order approaching
the Neel state by NMR Knight shift at the in-chain oxygen. The
nuclear spin-lattice relaxation rate offsets the $ \sim 1/T$
increase of the staggered spin susceptibility down to $10 K$ and
depends on frequency. The $q=0$ mode of spin transport is
ballistic and a diffusion-like contribution at $T << J $
\cite{Thurber} . The bulk susceptibility the measurements have
shown decreasing above $ \sim 10 K$ \cite{Motoyama} . The NMR
lineshape have been changed in a very specific way, giving rise to
"features", which have been explained by the chain edges and the
presence of mobile bond-defects. A "defect" is defined as a local
change in the magnetic bond coupling. As the translational
symmetry of the spin system is broken, a local alternating
magnetization develops around the defects in the direction of the
applied field. The density of bond-defects decreases linearly with
$T$. The source of the bond-defects might be related to the
interstitial excess oxygen. However , at the lowest temperature,
$T=20 K$, the density of bond-defects $\rho \simeq 2.8 \cdot
10^{-3}$ remains more than ten times larger than the concentration
of the residual spin-1/2 impurities $(\simeq 1.3 \cdot 10^{-4})$
after annealing and the observed $T$ dependence of $\rho$ is not a
constant \cite{Boucher} .

The concept of mobile "bond-defects" have been used for
explanation of the temperature dependence and the anisotropy of
the thermal conductivity below 50 K as the main source of spinon
scattering at low temperatures \cite{Sologubenko}. The calculated
mean distance between two neighboring bond defects, consistent
with the NMR data between 20 and 60 K. The results of the specific
heat measurements have been fit to the data above 4 K using the
approximation $C=\gamma T+\beta T^3 +\delta T^5 $. The linear in T
contribution is due to spinons excitations of a 1D Heisenberg
spin-1/2 antiferromagnetic system with exchange $J=2620 \pm 100$
which is $18 \%$  larger value than $J \simeq 2200$ deduced from
magnetic susceptibility data \cite{Motoyama} and NMR data
\cite{Boucher}. The Neel temperature $T_N$ dependents on the
measurement time and shifted to lower temperature with increasing
time. So Neel temperature is $T_N=5.4 K$ from neutron scattering
\cite{Kojima} , $4.15 K$ by $\mu SR$ \cite{Keren}  and $3.5 K $
from the anomalies of the specific heat \cite{Sologubenko}.
Possibly , here it is nonequilibrium process. Physical origin of
the bond-defects remains unclear also as abrupt change
suseptibility at different temperatures determined by NMR and by
external magnetic field.

In the paper we try to answer on these questions basing on
non-adiabatic interaction between spins . Numerical methods, such
as exact diagonalization and density-matrix renormalization group,
face the potential difficulty in dealing with a very large Hilbert
space of the phonons.  Here , we use Monte Carlo approaches
restricted to finite chains $L=100, 200$ but without any adiabatic
approximation and the truncation of the infinite phonon Hilbert
space. The method \cite{contin} is based on a path-integral
representation for discrete system in which we work directly in
the Euclidean time continuous. All the configuration update
procedures contain no small parameters . Being based on local
updates only, it allows to work with the grand canonical ensemble
and to calculate any
dynamical correlation function, expectation values.\\
{\bf 2. Model and method}

 The ratio intechain to intrachain exchange in $Sr_2CuO_4 $ is an extremely small
 $ \sim 10^{-3}$ and demonstrates a good one-dimensionality.  The
 structural fluctuation were only observed  along in chain at low
 temperature. As a result a spin-phonon interaction were
 considered along chain. The model Hamiltonian of an  spin-phonon system takes the form:
\begin{equation}\label{1}
  H=\sum_{i=1}^{L}[J+\alpha
(u_{i}-u_{i+1})][S_{i}^{z}S_{i+1}^{z}+(S_{i}^{+}S_{i+1}^{-}+S_{i}^{-}S_{i+1}^{+})/2]+M
\dot{u}_{i}^{2}/2+K(u_{i}-u_{i+1})^{2}/2.\;\;
\end{equation}
Here $S^{z,\pm} $ are a spin operator components associated with
the site i, $ J>0 $ is the usual antiferromagnetic exchange
integral, $\alpha$ is the spin-phonon coupling constant, $u_i$ is
the displacement in the z- direction, $M$ is the mass of the ion
and $K$ the spring constant. Using the quantum representation for
phonon operators $ b,b^{+} $, the Hamiltonian becomes
\begin{equation}\label{2}
H=\sum_{q}\sum_{i=1}^{L}[J+\alpha \sqrt{\frac{2\hbar }{M\omega _{0}}}\sqrt{%
\sin {\frac{q}{2}}}\cos {(q(i-0.5))}(b_{q}+b_{q}^{+})]
[S_{i}^{z}S_{i+1}^{z}+(S_{i}^{+}S_{i+1}^{-}+S_{i}^{-}S_{i+1}^{+})/2]\\
\end{equation}
$$
+\sum_{q}\hbar \Omega (q)b_{q}^{+}b_{q},\;\;\Omega (q)=2\sqrt{\frac{K}{M}}\sin {(%
\frac{q}{2})},\;\omega _{0}=2\sqrt{\frac{K}{M}},\;q=2\pi n/L,
n=1,2,...L
$$

The phonon frequency $(\omega_0)$, spin-phonon coupling
$(\alpha)$, energy $E$, temperature $T$ are normalized on the
exchange $J$ and $\hbar=1, M=1$. The temperature used in
calculation is $\beta=J/T=25$. Our system consists of the two
subsystem interacted. The elastic subsystem is described by
phonons with the number occupation $n_{ph}=0,1,2..$ and magnetic
subsystem is characterized by the number occupation $n_{m}=0,1$
and Pauli operators $a,a^+$  which coincide with $S^±$ spin
operators.  We start with the standard Green function of the
phonon in the momentum $q$- imarginary-time $\tau$ representation:

\begin{equation}\label{3}
 G(q,\tau )=<vac|b_{q}(\tau )b_{q}^{+}(0)|vac>,\;\tau \geq
0, b_{q}(\tau )=\exp {(H\tau )}b_{q}\exp {(-H\tau )}\; .
\end{equation}

$$
G(q,\tau
)=\sum_{\nu}|<\nu|b^{+}_q|vac>|^2\exp{[-(E_{\nu}(q)-E_{0})\tau]}
$$
where  ${|\nu>}$ is a complete set of the Hamiltonian $ H$ in the
sector of given $q$ , $ H|\nu(q)>=E_{\nu}(q)|\nu(q)>,\;
H|vac>=E_{0}|vac>, E_0=0$. Rewriting Eq.(3) as

\begin{equation}\label{4}
  G(q,\tau)=\int_0^{\infty} d\omega A(q,\omega) \exp{-(\omega
\tau)},\; \\
 A(q,\omega)=\sum_{\nu}
\delta(\omega-E_{\nu}(q))|<\nu|b^{+}_q|vac>|^2
\end{equation}
one defines the spectral function $A(q,\omega)$ . We calculate a
one-particle spin Green function

\begin{equation}\label{5}
  <a(\tau)a^{+}(0)>=\int_0^{\infty}d\omega \rho_t(\omega) \exp{-(\omega
\tau)},\\
\end{equation}
where  $\rho_t(\omega)$ -spectral density function. Correlation
functions are determined on the basis of a complete set of
eigenvector  of the Hamiltonian:
\begin{equation}\label{6}
  <O>=\frac{\sum_{\nu} <\nu_i|O|\nu_j>}{\sum_{\nu} <\nu_i|\nu_j>},
\end{equation}
where $O$ is the longitudinal spin-spin correlation function
$<S^z_iS^z_{i+h}>$ and the phonon density- density
$<n_{ph}(q)n_{ph}(q+p)>$. The distribution of  phonons number
$n_{ph}(q)$ and  the magnons number  $n_m(k) $
as a function of momentum are simulated.\\

{\bf 3. Results and discussion}

The upper boundary of  acoustic phonon band $\omega_0$ is in
direct proportion to Debye temperature of $\Theta_{Sr_2CuO_3}=441
\pm 10 K$ \cite{Sologubenko} in term of Debye approximation. Using
known data for $CuGeO_3$ , $\Theta_{CuGeO_3}=330 K$ \cite{Debye},
$\omega_{0,Ge} \simeq 1440 K$ \cite{Aplesnin} we estimate
$\omega_{0} \simeq 2100 K$ for $Sr_2CuO_3$. The ratio of
$\omega_{0}/J \simeq 0.95$ will be used for calculation of
dynamical characteristics  . The spectral density of triplet
excitations are shown in Fig.1. At the critical value of
spin-phonon coupling $\alpha_{c1} $ the spectral density $\rho_t
=0$ for $E<\Delta$, were $\Delta$ is the gap in the triplet
excitation spectrum. The gaps energies presented in inset of Fig.1
fit satisfactory on the straight line
$\Delta=-0.17(2)+1.16(5)\alpha$ in the range of $\alpha_{c1}<
\alpha < \alpha_{c2} $ with $\alpha_{c1}=0.15(1)$.

For small spin-phonon coupling $\alpha < \alpha_{c1} $ the
composite quasiparticle consisted spinon and phonon are formed.
The inelastic interaction between spinon with momentum $k$ and
phonon with $q$ lead to new quasiparticle with energy $E=E_q+E_k$
and momentum $k+q$. It follow from the distribution of the phonon
number versus momentum as plotted in Fig.2a. The density of spinon
excitations exhibits the singularity at the $E(k=\pi/2)$ and
phonon excitations at the $E(q=\pi)$ and summered momentum agrees
with MC result $N_{ph}(q\simeq 4.5) \neq 0 $. Quasiparticles with
 $q-k$ are absent . It may be attributed to the repulsion
spin-phonon quasiparticle with momentum $q-k$ from spinon with the
same energy $E=E_k-E_q$. Whereas the density of spinon excitations
tends to zero for energies $E=E_q+E_k$. The number of spin-phonon
quasiparticles is increased against $\alpha$ as illustrated in
Fig.2 and the bound states of spin-phonon quasiparticls are formed
at $\alpha >\alpha_{c1} $ that are similar to bipolaron with the
symbolic kind $<b^+_{\textbf q-\sum_i^{n_1} k_i}b^+_{\textbf
q+p-\sum_i^{n_2} k_i} \prod_i^{n_1+n_2} S^z_{\textbf k_i }> $.

The correlation function of the phonon numbers
$<N_{ph}(k)N_{ph}(k+q)>$, presented in Fig.3a , is nonzero at the
certain   wavenumber  $q \sim 0.1$ and $q \sim \pi/3$
  for   $\alpha \simeq \alpha_{c1}$ and for all $q$ at
   the condition  $\alpha > \alpha_{c2}$. The elastic strains
   cause nonuniform distribution of exchange and spin density that
   confirms spin-spin correlation function. The difference between
   spin-spin correlation function simulated in chain with
   spin-phonon coupling and without it is shown in Fig.3b. Small
   oscillations are observed for $\alpha < \alpha_{c2} $. Cycle of
the oscillations $\delta r =9$ well agrees with cycle of
oscillations of the magnons number distribution function $\delta
k=0.36, 0.86; \delta r=\pi/\delta k \simeq 9 $ plotted in Fig.4.
The distribution of phonon numbers becomes continuum at the
$\alpha
> \alpha_{c2}$ with one sharp maximum at the wave vector of
structure being in the range of  $\pi < Q < 2 \pi$ as plotted in
Fig.2b. The value $\alpha_{c2}=0.4 \pm 0.02$ is similar to a
critical concentration when the dressed phonons effectively
"percolate " along the three low-lying bands. The correlation
function $<N_{ph}(k)N_{ph}(k+q)>$ is not equal to zero for all
momentums as shown in Fig.3.

The transition from one state with localized phonons to
delocalized is like to Anderson's transition in disordered
systems. The spin density in chain allocates random and nonuniform
that follow from the normalized spin-spin correlation function
shown in Fig.3b. The estimated values fit by exponential
approximation  $\mid<S^z_0S^z_r>-<S^z_0S^z_r>_{AF} \mid \sim A
\exp {(-r/\xi)}$, where $\xi$ is a correlation radius of
inhomogeneity. The magnons number  distribution function exhibits
also the random oscillations (Fig.4c). The transition to
delocalized phonons is accompanied by strong increasing of the
mean-square amplitude of ions vibration as presented in Fig.5b. At
$\alpha=\alpha_{c1} $ the function $<u^2>(\alpha)$ reveals also
 a some peculiarity. The average number of phonons , plotted in
 Fig.5a , increases vs. $\alpha$ according to the power law
$$N_{av}=0,0012
(\frac{\alpha}{\alpha_{c2}})^{1.75(6)} (7)
$$

From  the spectral density of phonons shown in Fig.6 and spin
excitations it may be stated that the coherent nonlinear bound
excitations of phonon and magnon exist at $\alpha >\alpha_{c1} $.
It may be interpreted as a standing vibration with the spin kink
in sites. The  disjoint  vibrations are performed at the condition
that the wave length is changed in two times $l/a=2; 4; 8; 16;
...$ c $k=\pi/l, E=w_0 sin(k/2)$. These  estimates of energy are
in good agreement with Monte Carlo results up to $l/a=32$ because
the temperature  fluctuations  $(T=0.04)$ cut the kinks
interaction radius. \\
Now we will apply our results to estimate spin-phonon coupling for
the one-dimensional spin system $Sr_2CuO_3$ with intrachange
$J=2200 K, \omega_0=2100 K$. The abrupt change of the bulk
susceptibility at $T \simeq 10 K$ is associated with gap in the
triplet excitation spectrum. Using typical ratio between gap and
critical temperature, for example, for $CuGeO_3,\; \Delta/T_c
\simeq 1.8$ we estimate $\alpha=0.154$ from eq. $0.008=-0.17+1.16
\alpha$. The average phonons number is $N_{ph}=2.2\cdot 10^{-4}$.
These phonons are formed bound states with energy about $ \sim
0.55 \; eV \;, 0.76 \; eV,\; 1\; eV$ as followed from the spectral
phonon density plotted in Fig.6 because maximum phonon energy with
$n_q=1, E=0.95J \sim 0.18 \;eV$. This energy of elastic
fluctuation is compared to the value of the experimental charge
transfer gap $E_g=1.5 \pm 0.3 eV$ derived from He I ultraviolet
photoemission combined with bremsstrahlung isochromat spectroscopy
\cite{Maiti} and the on-site Coulomb interaction  on oxygen site
$U_p=4.4 eV$ \cite{Neudert} . The elastic fluctuation lead to
change of the electron density on oxygen site
  and discontinuity of
antiferromagnetic exchange that causes formation of the
paramagnetic spin $1.3 \cdot 10^{-4}$ observed at low temperature
$ \sim 4 K$.

There are spin-phonon quasiparticles with energy $E_{s,ph}=1040
cm^{-1}, 562 cm^{-1},  287 cm^{-1}$,

$144 cm^{-1}, 72 cm^{-1}, 36 cm^{-1}, 18 cm^{-1}$. The optic
phonon mode observed at $\omega_1=550 cm^{-1}$ corresponds to
stretching mode and mode with $\omega_2=569 cm^{-1}$ is attributed
to spin-phonon quasiparticle with $k=\pi/4, \omega^{MC}=562
cm^{-1}$. The relaxation time (or lifetime ) of the bound
spin-phonon quasiparticle is decreased at heating and at $T \sim
E^{MC} \sim 800 K$ it is decomposed that qualitatively agrees with
experimental data. The phonon width of optical conductivity
spectrum is increased from $\gamma=17.4 cm^{-1}, T=10K$ to
$\gamma=35.1 cm^{-1}, T=300K$ and the strength of the oscillator
is decreased from $S=0.09 cm^{-1} eV, T=10 K$ to $S=0.06 cm^{-1}
eV, T=300 K$. The phonon width and the strength of the oscillator
of mode $\omega_1=550 cm^{-1}$ don't change in the
range $10 K < T < 300 K$ \cite{Lee}. \\
 The bond defects observed by NMR and heat transport arise from
formation of the standing vibration sites. The density of bond
defects $\rho \simeq 1/l$ is in direct proportional to wavevector
of the standing vibration $k=\pi/l$ and estimated density is
satisfactory fit with experimental data at $T=20 K, \rho^{MC}
\simeq 5 \cdot 10^{-3}, \rho^{ex} \simeq 2.8 \cdot 10^{-3}$. If
discrete spectrum of spin-phonon quasiparticles map to continuum
the temperature dependence $\rho$ may be presented as $\rho \simeq
2 T/\pi \omega_0 $ which well agree with NMR data \cite{Boucher}.
It is also explained the linear dependence of the specific heat at
low temperatures. Interaction between  loosely coupled phonon and
spinon modes in the center of Brillouin band may be presented as
$(k-\omega/v_{ph})(k-\omega/v_{sp})=\tilde{\alpha}^2$, where
$v_{ph}=\omega_0/2, v_{sp}=\pi/2 J,
\tilde{\alpha}=\alpha/\sqrt{\omega_0}$. It lead to repulse of
these branches and formation of gap in the spinon spectrum with
$\Delta(k=0)/J=\alpha \sqrt{\pi}\ /2\sqrt{J} \simeq 0.019$. This
gap $\Delta(k=0)=43 K$ causes abrupt decrease of the dynamical
spin susceptibility for $k=0$ at $T \sim 33 K$. The $k=0$ mode of
spin transport has a diffusion like contribution for $T<< J$ as a
result of the fact that the finite density of bound spin-phonon
quasiparticles exists  at $E=0$ .

 At low temperature the spin liquid with non-linear topological excitations
 having $S=1/2$  is freezed
 that lead to small magnetic moment on site and elastic neutrons scattering.
 So in compound $CuGeO_3$ having transition singlet-triplet at $T=14
 K$ a long range order is observed at low temperatures when ions
 $Cu$ or $Ge$ are displaced by $Mg$ or $Sr$. Breaking a exchange
 between the nearest neighbors results in a finite spin density near breaking.
 Between the generated magnetic moments the interaction is existed per phonon field
 in 3D dimensional space. The critical field may be estimated by
 mean field (MF) approximation $T_N=2Sc (J_0+2J_1)$, where $c-$
 impurities concentration, $S=1/2$ , $J_0$ and $J_1$ is
 accordingly intra- and interchain exchange in simple cubic
 lattice. The calculated critical temperatures for
 $Cu_{1-x}Mg_xGeO_3$ are in well agreement with experimental data
 $T_N^{MF}=2.4 K, T_N^{ex}=2.5 K, x=0.016;\; T_N^{MF}=3.2 K, T_N^{ex}=3 K,
 x=0.0216$ \cite{Takeya } . Using similar evaluation for $Sr_2CuO_3$ with
 concentration of bond defect at $T \simeq 6 K, c \simeq 2 \cdot 10^{-3}$
  we  determined the critical temperature $T_N^{MF} \simeq 4.5 K$
  which is in well agreement with experiment $3.5 K < T_N^{ex} <
  5.4 K$. Scatter of temperatures may be due to dependence of
  freezing bond defects from frequency. The average distance
  between bond defects is $ \sim 2 \cdot 10^{3} A$ and
   the  periodic local alternating magnetization exists which is observed
  by NMR at low temperatures. Mean square displacement is $<U^2> \sim 10^{-6}$
  that may give rise the structural deformation along chain $\delta a \sim \sqrt{<U^2>} \sim 10^{-3}
   A$ for $a=3.9 A, M \simeq 10^{-22} g$.

  In summary, the spectrum of spinon excitations along chain is
  asymmetric relative wavenumber $k=\pi/2$ with two gaps $\Delta(k=0)=43
  K $ and $\Delta(k=\pi)=18 K$ that account for  steeply
  decrease of the bulk susceptibility and the dynamical spin susceptibility for
$k=0$ at the temperatures $T \simeq 10K, T \simeq 33 K$. The
spectrum of bound spin-phonon excitations is due to bond defects
the density of which  is linearly increased against temperature.
At low temperature $T \sim 5 K $ the non-linear spin-phonon
excitations are freezed  on the breaking exchange and induce
magnetic moment on site with correlation radius of
antiferromagnetic ordering $1200 - 2000 A$. The magnetic
properties of $Sr_2CuO_3$ are similar to $CuGeO_3$ with impurities
having two transitions on temperature. The paramagnetic
contribution in the bulk susceptibility at $T< 4 K$ result from a
structural defects on oxygen ions due to spin-phonon coupling.

   The author are grateful to V.A. Kashurnikov, A.S. Mishchenko,
   N.B. Prokof'ev, B.S. Svistunov for useful assistance and
   discussions of Monte Carlo method in the continuous time.

\newpage
Captions to Aplesnin paper  " Quantum spin liquid in $Sr_2CuO_3$
induced by  Acoustic Phonons "

 Fig.1  The spectral density of the one-particle excitations with $\omega_0=0.95,
 \alpha=0.05 (1), 0.15(2) \; (a)$; $ \alpha=0.2 (1), 0.4 (2) (b) $. In insert,
   the gap energy $\Delta$ in the
 excitation spectrum as a function of spin-phonon coupling .

Fig.2 Distribution of phonon numbers versus momentum for $ \alpha=
0.05 (1), 0.2(2)\; (a)$; $ \alpha=0.4 (1), 0.6 (2) (b) . $

Fig.3 The normalized correlation function of the phonon numbers
versus wavenumber for $\alpha=0.2 (1), 0.4 (2), 0.6 (3)$$ (a)$ and
the spin-spin correlation function as a function of distance for
$\alpha=0.2 (1), 0.6 (2)$ $(b)$ . Solid line is shown the
exponential dependence $Aexp{(-r/\xi)}, A=0.008(8), \xi=9.1(9)$.

Fig.4 Distribution of magnon numbers versus momentum for $ \alpha=
0.05 ,0 \leq k \leq \pi \;(a)$; $ \alpha=0.2 , 0 < k < \pi \;(b);
\alpha=0.6 , 0 < k < \pi \;(c). $

Fig.5 The average phonons number $<N_{ph,av}> (a) $ and the
mean-square displacement of ions $<u^2> (b)$ versus
$\alpha/\alpha_{c2}$.

Fig.6 The normalized spectral density of phonon excitations
  $ \rho_{ph}/\rho_{ph,max}$ versus energy  for the
spin-phonon  coupling  $  \alpha=0.25 (a); \; \alpha=0.4 (b);\;
\alpha=0.7 (c)$.

$\vert$%


\newpage


\begin{thebibliography}{14}

\bibitem{Abrashev}  Abrashev M.V.,  Litvinchuk A.P.,  Thomsen C.,  Popov V.N. 1997 Phys. Rev. B
{\bf 55} R8638 .

\bibitem{Lee} Lee Y.S. , Noh T.W. and et. al. 2000
Phys. Rev. B {\bf 62} 5285.

\bibitem{Thurber} Thurber K.R., A.W. Hunt and et . al. 2002 cond-mat/ 0108272.

\bibitem{Motoyama} Motoyama N., Eisaki H. and Uchida S. 1996
Phys. Rev. Lett. {\bf 76} 3212.

\bibitem{Boucher} Boucher J.P., Takigawa M. 2000, Phys. Rev. B {\bf 62} 367.

\bibitem{Sologubenko} Sologubenko A.V., Felder E.
and et. al. 2000 Phys. Rev. B {\bf 62} R6108.

\bibitem{Kojima} Kojima K.M. et. al. 1997 Phys. Rev. Lett. {\bf 78} 1787.

\bibitem{Keren} Keren A. et. al. 1993 Phys. Rev.
B {\bf 48} 12 926.


\bibitem{contin}  Beard B.B., and  Wiese U.J. 1996 Phys. Rev. Lett. {\bf
58}, 5130 ; Prokof'ev  N.V.,  Svistunov B.V. 1998 Phys. Rev. Lett.
{\bf 81}, 2514 ; Prokof'ev  N.V. ,  Svistunov B.V.,  Tupitsin I.S.
1998  Zh. Eksp. Teor. Fiz.  {\bf 114}, 570 , [ JETP {\bf 87 }, 310
(1998)];  Mishchenko A.S.,  Prokof'ev N.V.,  Sakamoto A.,
Svistunov B.V. 2000 Phys. Rev. B {\bf 62} 6713 .

\bibitem{Debye} Lasjaunias J.C. ,  Monceau P.,  Remenyi G. and et.
al. 1997 Solid State Commun. {\bf 101} 677 .

\bibitem{Aplesnin} Aplesnin S.S. 2002 submitted in Phys. Rev. Lett.,
cond-mat/0207118.


\bibitem{Maiti} Maiti K., Sarma D.D. and et. al. 1997 Europhys. Lett. {\bf
37 }  359.

\bibitem{Neudert} Neudert R., Drechsler S.L. and et. al. 2000 Phys. Rev.
B {\bf 62} 10752.

\bibitem{Takeya } Takeya J., Tsukada I. and et. al. 2000 Phys. Rev. B
 {\bf 62} R9260.


\end{thebibliography}
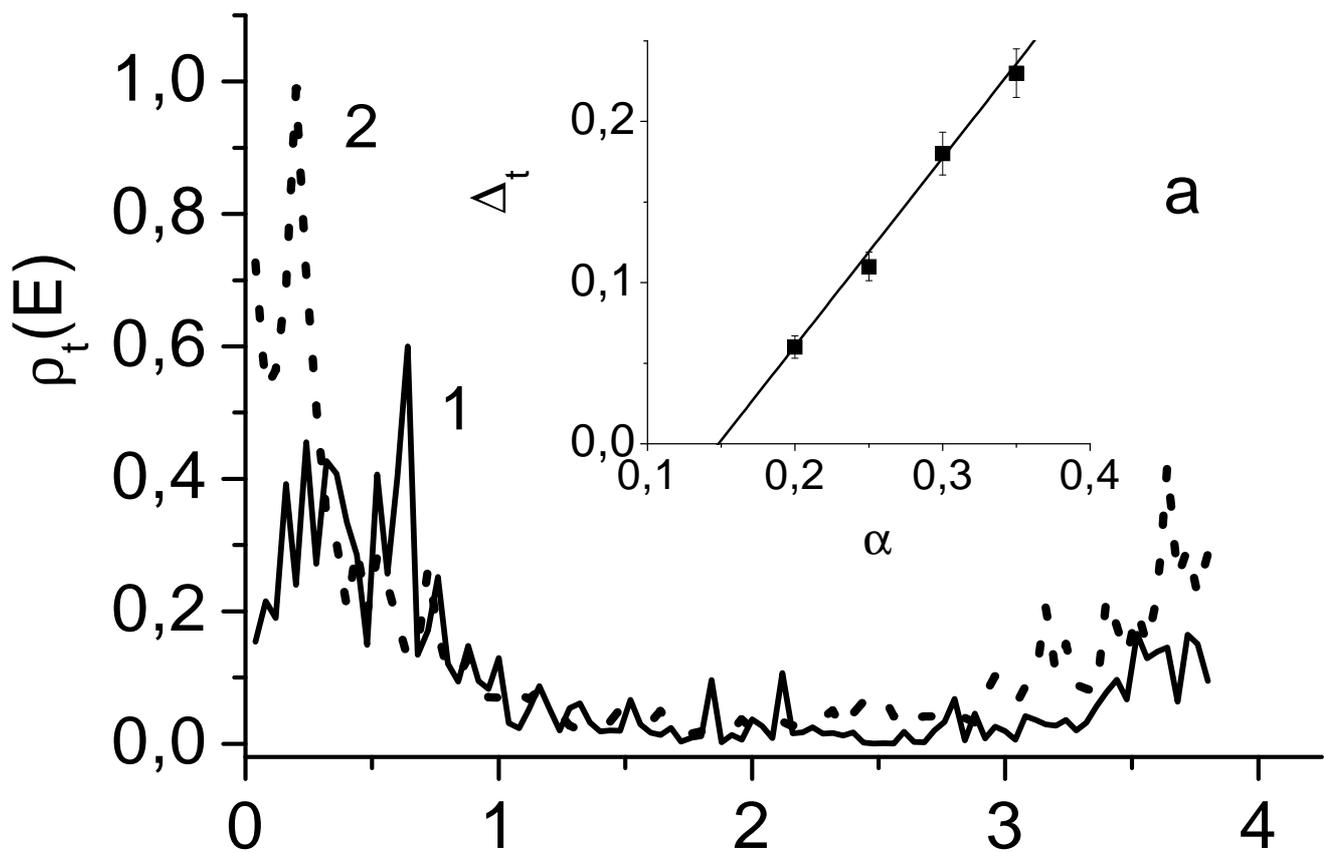
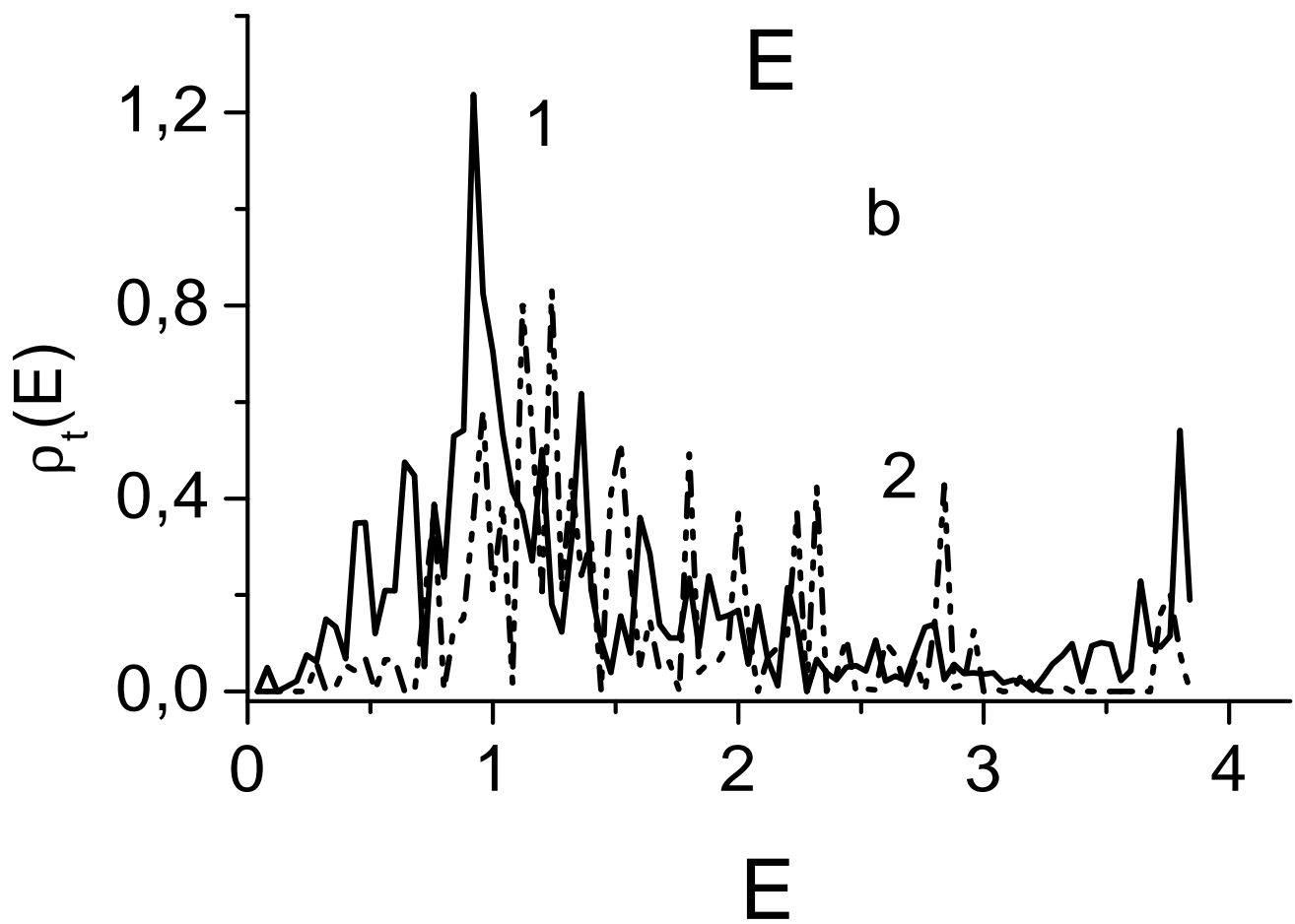
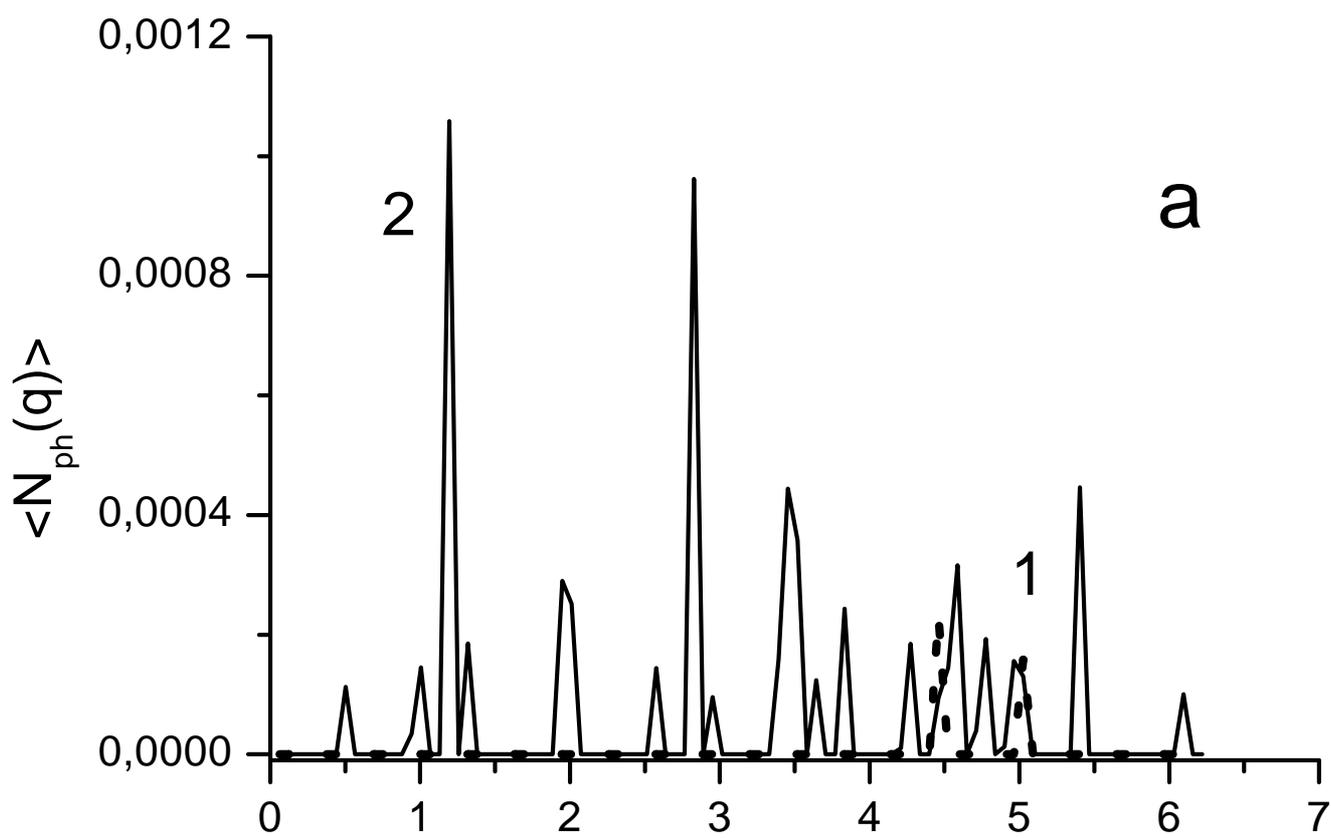
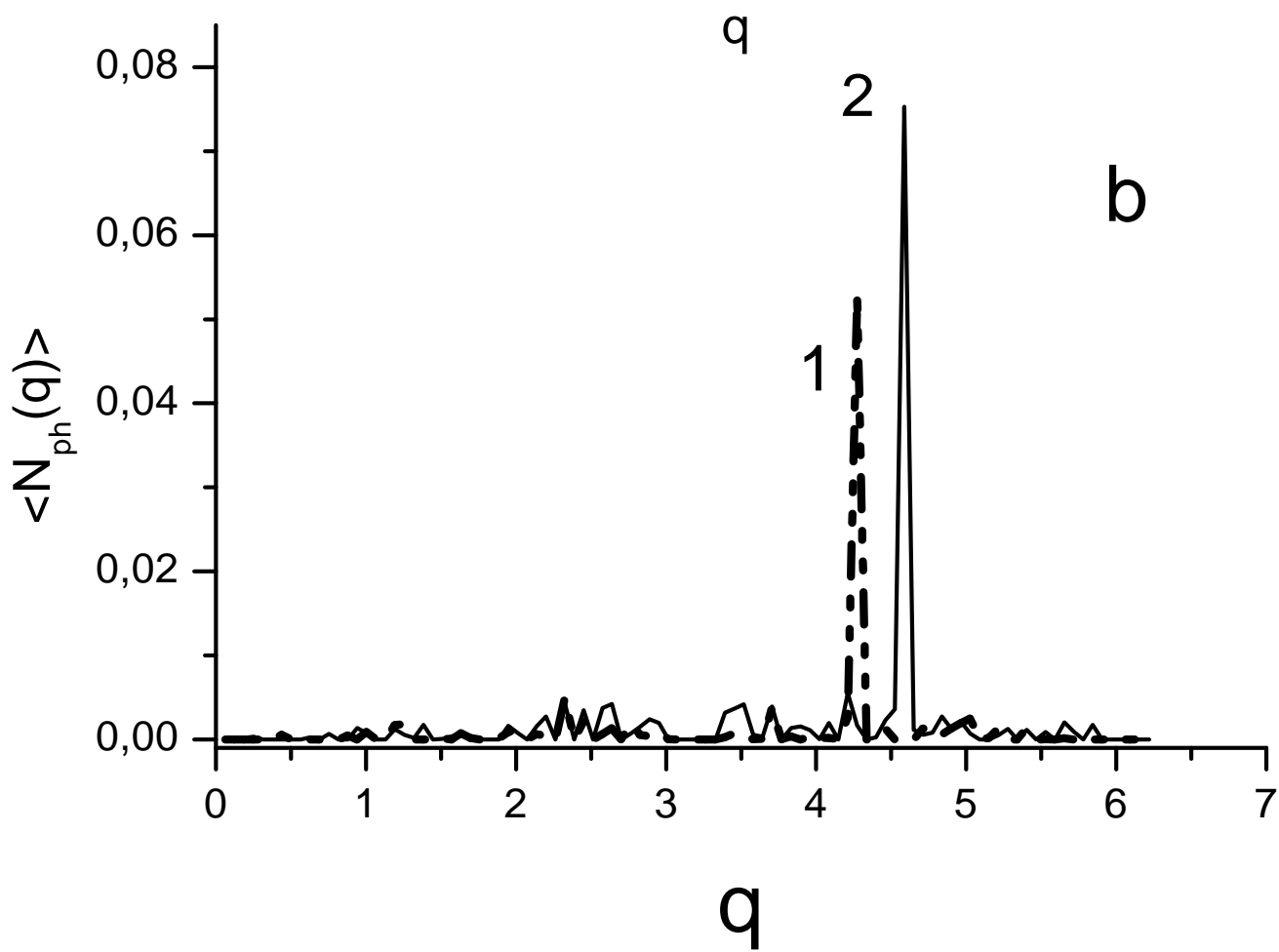
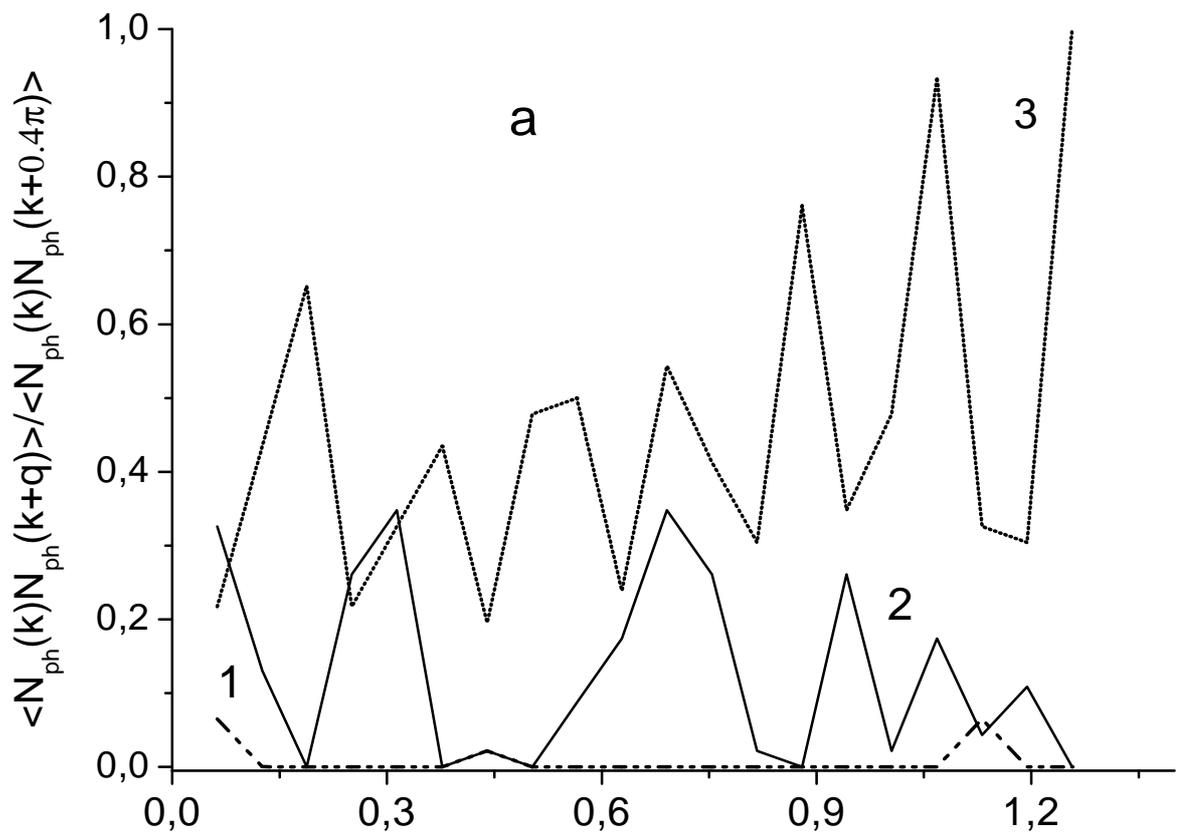
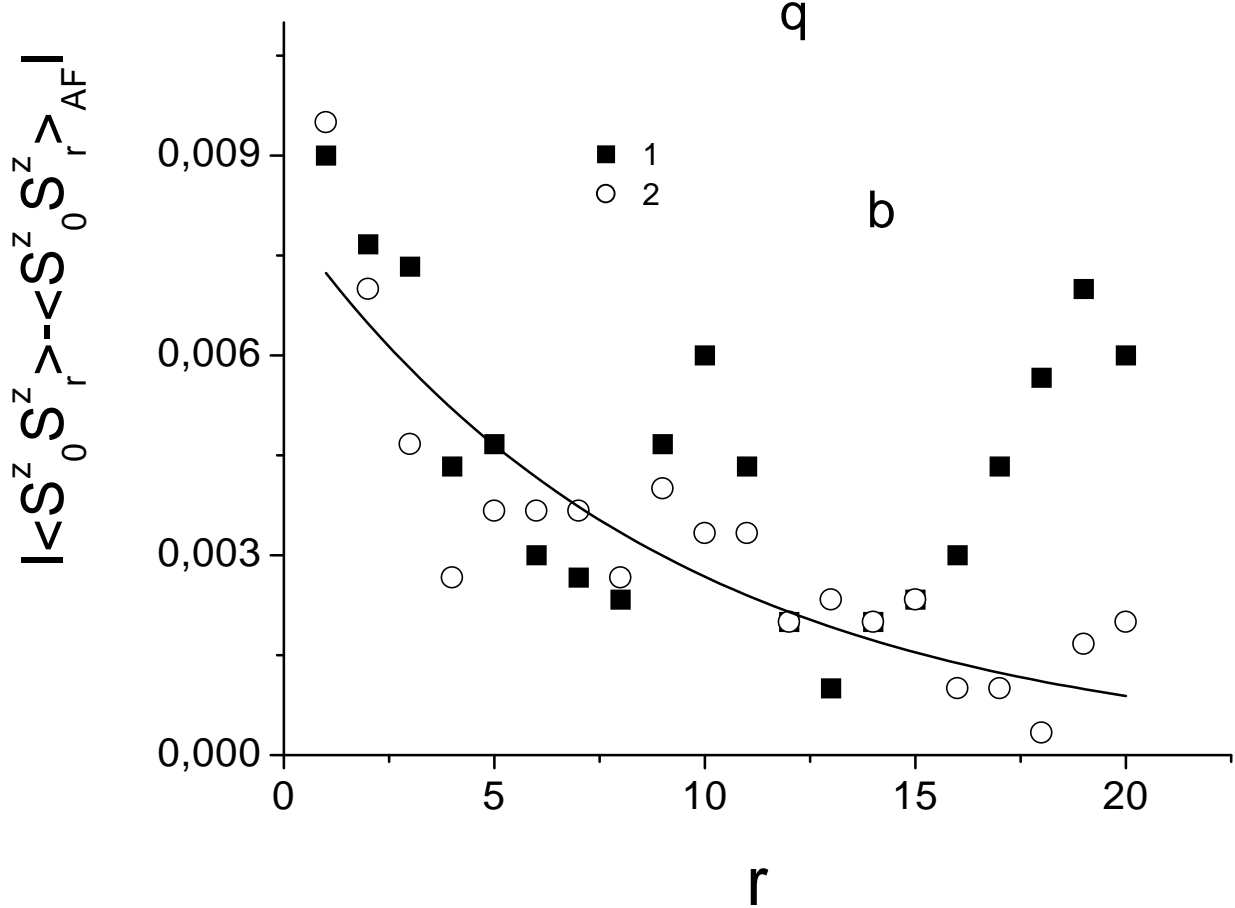
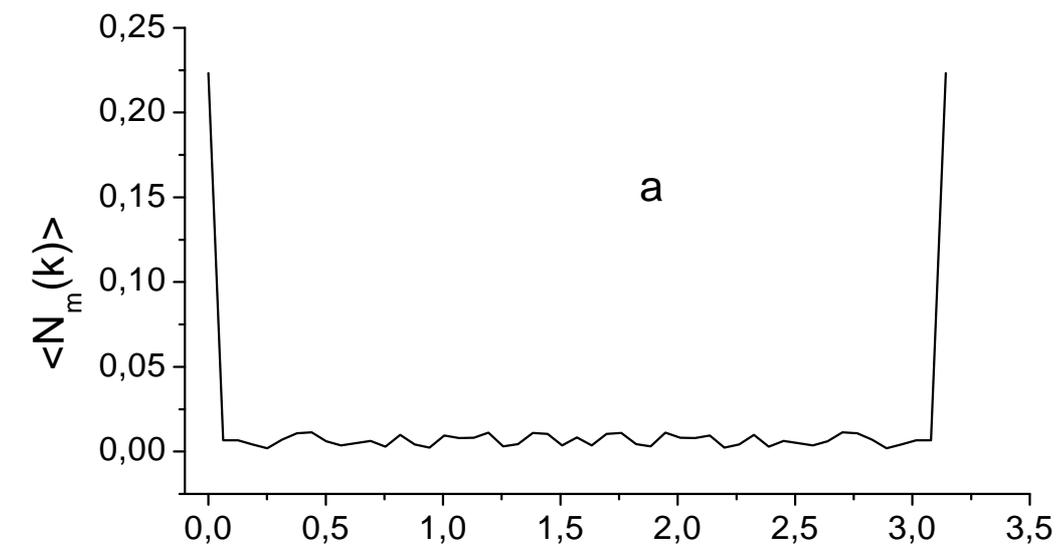
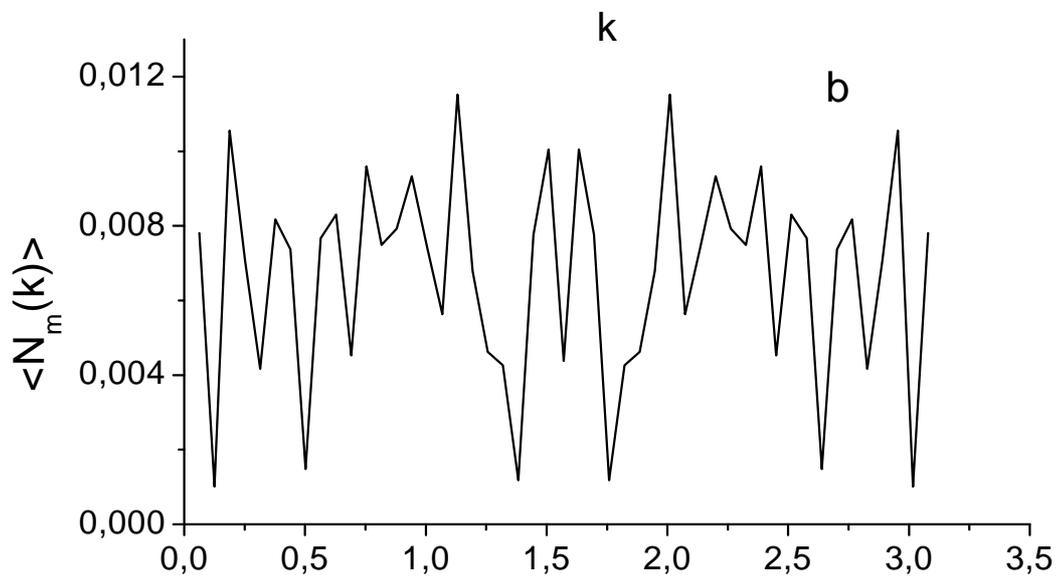
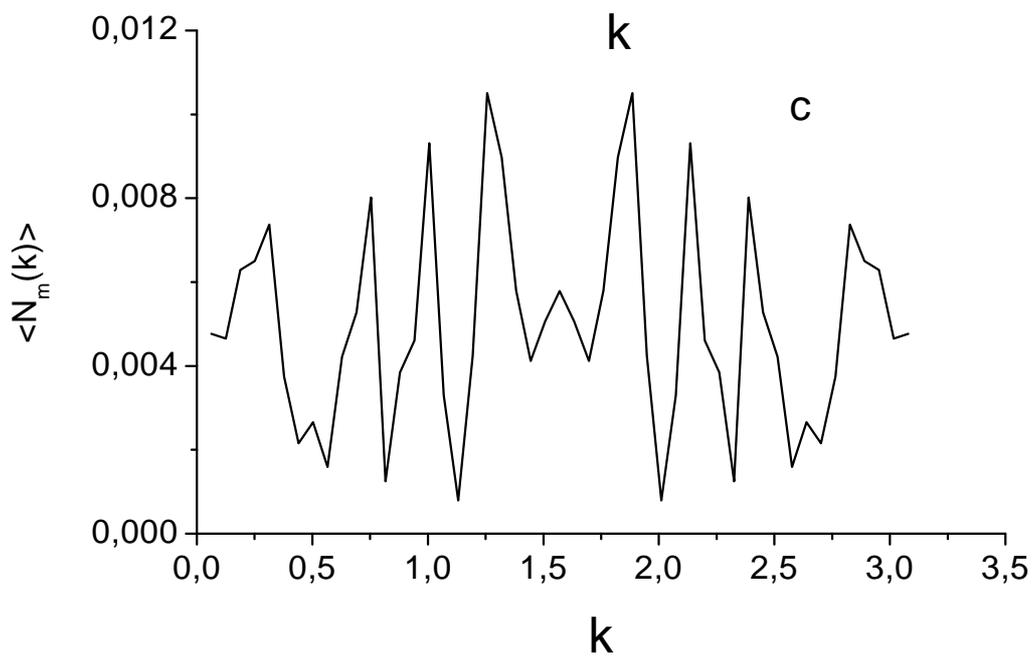
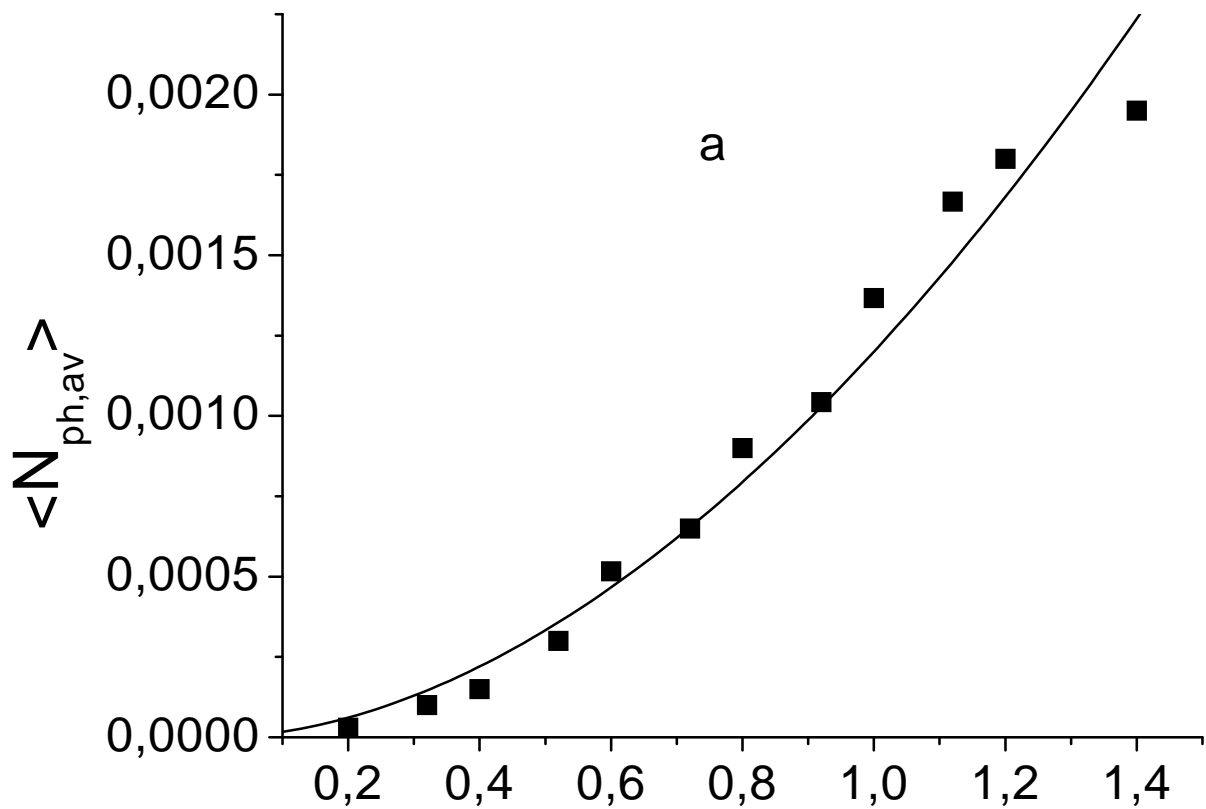
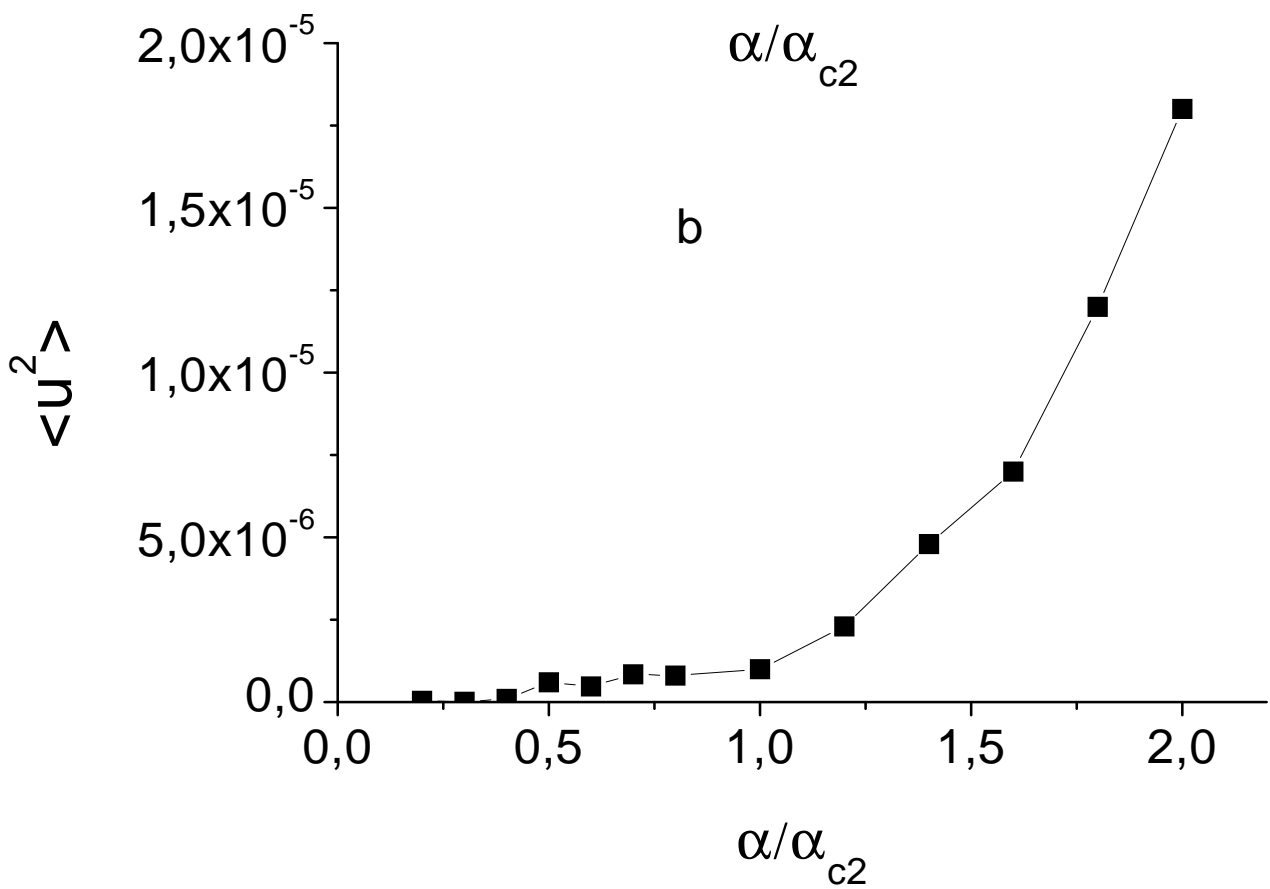
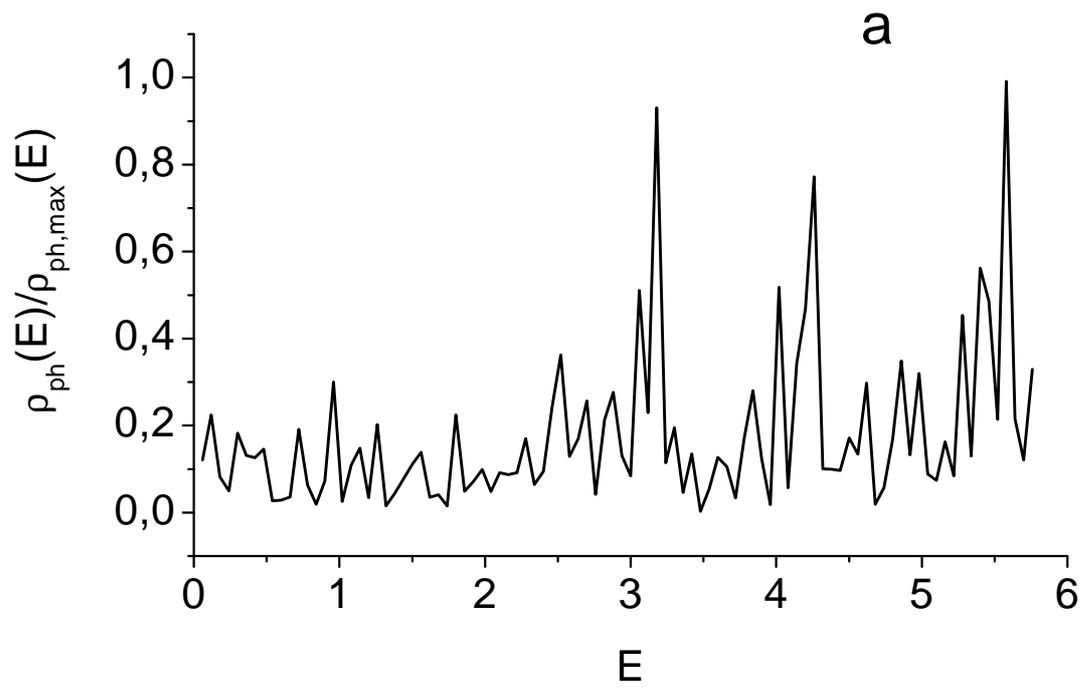
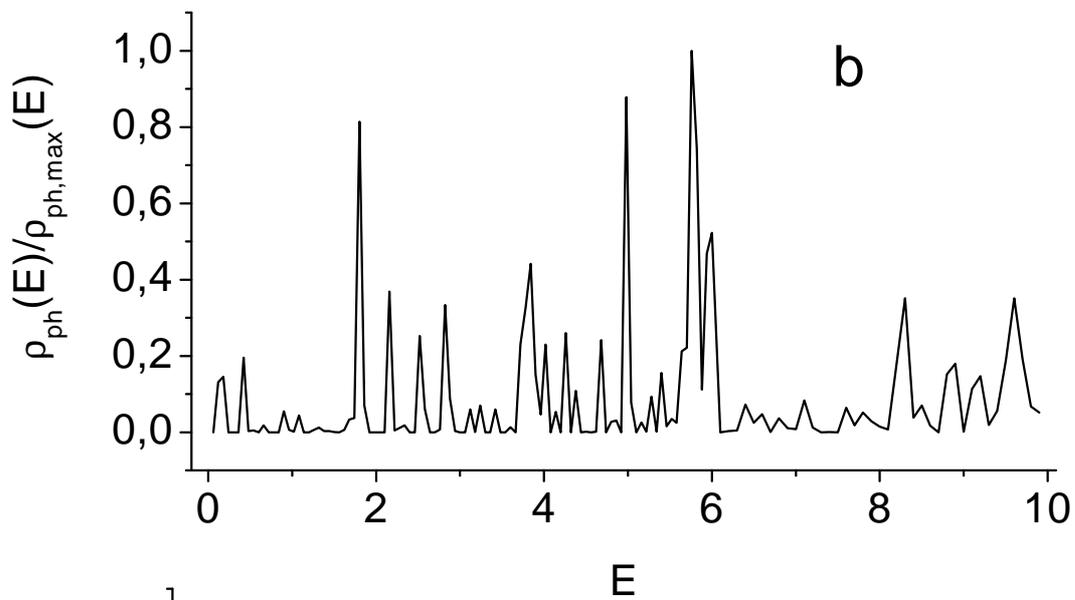
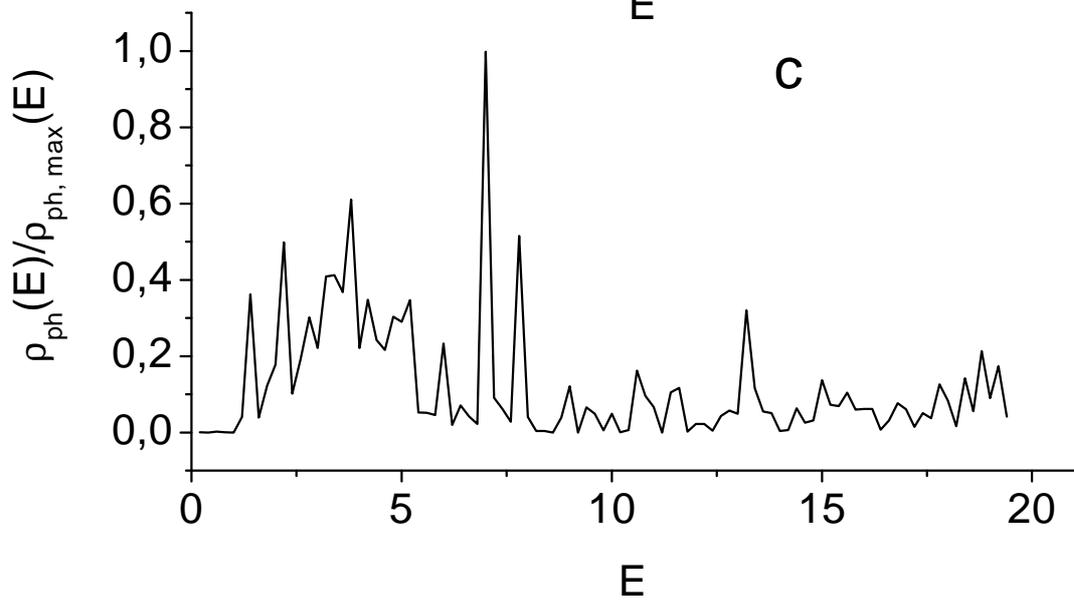
\end{document}